\documentstyle[12pt,epsf]{article}
\begin{document}
%%%%%%%%%%%%%%%
\catcode`@=11
% Redefine caption to put text and formulas in smaller font
\long\def\@caption#1[#2]#3{\par\addcontentsline{\csname
  ext@#1\endcsname}{#1}{\protect\numberline{\csname
  the#1\endcsname}{\ignorespaces #2}}\begingroup
    \small
    \@parboxrestore
    \@makecaption{\csname fnum@#1\endcsname}{\ignorespaces #3}\par
  \endgroup}
\catcode`@=12
%%%%%%%%%%%%%%%%%%%%%%%%%%%%%%%%%%%%%%%%%%%%%%%%%%%%%%%%%%%%
\newcommand{\newc}{\newcommand}
\newc{\gsim}{\lower.7ex\hbox{$\;\stackrel{\textstyle>}{\sim}\;$}}
\newc{\lsim}{\lower.7ex\hbox{$\;\stackrel{\textstyle<}{\sim}\;$}}
\newc{\gev}{\,{\rm GeV}}
\newc{\mev}{\,{\rm MeV}}
\newc{\sw}{\sin\theta_W}
\newc{\cw}{\cos\theta_W}
\newc{\swsq}{\sin^2\theta_W}
\newc{\cwsq}{\cos^2\theta_W}
\newc{\five}{{{\bf5}}}
\newc{\ten}{{{\bf10}}}
\newc{\sixteen}{{{\bf16}}}
\newc{\fpf}{{{\bf\bar5}+{\bf5}}}
\newc{\tpt}{{{\bf\overline{10}}+{\bf10}}}
\newc{\sps}{{{\bf\overline{16}}+{\bf16}}}
\newc{\nf}{n_5}
\newc{\nt}{n_{10}}
\newc{\mx}{M_X}
\newc{\mstr}{M_{string}}
\newc{\mpl}{M_{Pl}}
\newc{\mz}{M_Z}
\newc{\mw}{M_W}
\newc{\mnew}{M_n}
\newc{\mhalf}{M_{1/2}}
\newc{\gstr}{g_{string}}
\newc{\gx}{g_X}
\newc{\alx}{\alpha_X}
\newc{\alsz}{\alpha_3(M_Z)}
\newc{\ytil}{{\tilde y}}
\newc{\nieff}{n_{\it i,eff}}
\newc{\nfeff}{n_{5,{\it eff}}}
\newc{\nteff}{n_{10,{\it eff}}}
\def\vev#1{\left\langle #1 \right\rangle}
\def\fracs#1#2{{\textstyle{#1\over#2}}}
\def\fract#1#2{{\scriptstyle #1/#2}}
%     curly letters
\def\CAG{{\cal A/\cal G}}   %curly letters
\def\CA{{\cal A}} \def\CB{{\cal B}} \def\CC{{\cal C}} \def\CD{{\cal D}}
\def\CE{{\cal E}} \def\CF{{\cal F}} \def\CG{{\cal G}} \def\CH{{\cal H}}
\def\CI{{\cal I}} \def\CJ{{\cal J}} \def\CK{{\cal K}} \def\CL{{\cal L}}
\def\CM{{\cal M}} \def\CN{{\cal N}} \def\CO{{\cal O}} \def\CP{{\cal P}}
\def\CQ{{\cal Q}} \def\CR{{\cal R}} \def\CS{{\cal S}} \def\CT{{\cal T}}
\def\CU{{\cal U}} \def\CV{{\cal V}} \def\CW{{\cal W}} \def\CX{{\cal X}}
\def\CY{{\cal Y}} \def\CZ{{\cal Z}}
%%%%%%%%%%%%%%%%%% Reference Defs %%%%%%%%%%%%%%%%%%
\def\NPB#1#2#3{Nucl. Phys. {\bf B#1} (19#2) #3}
\def\PLB#1#2#3{Phys. Lett. {\bf B#1} (19#2) #3}
\def\PLBold#1#2#3{Phys. Lett. {\bf#1B} (19#2) #3}
\def\PRD#1#2#3{Phys. Rev. {\bf D#1} (19#2) #3}
\def\PRL#1#2#3{Phys. Rev. Lett. {\bf#1} (19#2) #3}
\def\PRT#1#2#3{Phys. Rep. {\bf#1} (19#2) #3}
\def\ARAA#1#2#3{Ann. Rev. Astron. Astrophys. {\bf#1} (19#2) #3}
\def\ARNP#1#2#3{Ann. Rev. Nucl. Part. Sci. {\bf#1} (19#2) #3}
\def\MODA#1#2#3{Mod. Phys. Lett. {\bf A#1} (19#2) #3}
\def\ZPC#1#2#3{Zeit. f\"ur Physik {\bf C#1} (19#2) #3}
\def\APJ#1#2#3{Ap. J. {\bf#1} (19#2) #3}
%%%%%%%%%%%%%%%%%%%%%%%%%%%%%%%%%%%%%%%%%%%%%%%%%%%%%%%%%%%%%
\def\beq{\begin{eqnarray}}
\def\eeq{\end{eqnarray}}
\def\bea{\begin{eqnarray}}
\def\eea{\end{eqnarray}}
\newc{\ie}{{\it i.e.}}          \newc{\etal}{{\it et al}}
\newc{\eg}{{\it e.g.}}          \newc{\etc}{{\it etc}}
%%%%%%%%%%%%%%%%%%%%%%%%%%%%%%%%%%%%%%%%%%%%%%%%%%%%%%%%

\let\al=\alpha
\let\be=\beta
\let\ga=\gamma
\let\Ga=\Gamma
\let\de=\delta
\let\De=\Delta
\let\ep=\varepsilon
\let\ze=\zeta
\let\ka=\kappa
\let\la=\lambda
\let\La=\Lambda
\let\del=\nabla
\let\si=\sigma
\let\Si=\Sigma
\let\th=\theta
\let\Up=\Upsilon
\let\om=\omega
\let\Om=\Omega
\def\imp{~\Rightarrow}
\let\p=\partial
\let\<=\langle
\let\>=\rangle
\let\ad=\dagger
\let\txt=\textstyle
\let\h=\hbox
\let\+=\uparrow
\let\-=\downarrow
\def\L{{\cal L}}
\def\dot{\!\cdot\!}
\def\tr{{\rm tr}}
\def\vfilll{\vskip 0pt plus 1filll}
\def\ee#1{\times 10^{#1} }

% Style-sensitive Poor-Man's-Bold command, produces bold greek letters.
% Usage $ ... \pmb\gamma ... $
% Adapted from TeXbook p386 (\pmb) and p360 (\mathpallette)
%
\newdimen\pmboffset
\pmboffset 0.022em
\def\oldpmb#1{\setbox0=\hbox{#1}%
 \copy0\kern-\wd0
 \kern\pmboffset\raise 1.732\pmboffset\copy0\kern-\wd0
 \kern\pmboffset\box0}
\def\pmb#1{\mathchoice{\oldpmb{$\displaystyle#1$}}{\oldpmb{$\textstyle#1$}}
        {\oldpmb{$\scriptstyle#1$}}{\oldpmb{$\scriptscriptstyle#1$}}}
% End of pmb

\let\ol=\overline
\def\nl{\hfil\break}
\pretolerance=10000  %No hyphens

%%%%%%%%%%%%%%%%%%%%%%%%%%%%%%%%%%%%%%%%%%%%%%%%%%%%%%%%%%%%%%%%%%%%%%%%%%%%

\begin{titlepage}
\begin{flushright}
{IASSNS-HEP-96-42\\
hep-ph/9609480\\
September 1996\\}
\end{flushright}
\vskip 2cm
\begin{center}
{\Large\bf Low-Energy Signatures of Semi-perturbative
Unification\footnote{Research supported by Dept.~of Energy contract
\#DE-FG02-90ER40542, and by the W.M.~Keck and Monell Foundations.
Email addresses: kolda@sns.ias.edu, jmr@sns.ias.edu}}
\vskip 1cm
{Christopher Kolda and John March-Russell\\}
\vskip 5pt
{\sl School of Natural Sciences\\
Institute for Advanced Study\\
Princeton, NJ 08540}
\end{center}
\vskip .5cm
\begin{abstract}
We consider the low-energy signatures of, and high-energy motivations
for, scenarios of semi-perturbative gauge coupling unification.
Such scenarios can leave striking imprints on the low-energy sparticle
spectrum, including novel gaugino mass ratios (including
$M_2/M_1\approx1$), substantial compression of the intra-generational
squark-to-slepton mass ratios, and an overall lifting of scalar masses
relative to the gauginos. We also demonstrate that the unification scale
can be raised to $M_X \approx 4\times10^{17}~\gev$ while still in the
perturbative regime -- close to the 1-loop heterotic string scale. 
We employ a 3-loop calculation of the running of the gauge couplings
as a test of the perturbativity of the high-scale theory.
\end{abstract}
\end{titlepage}
\setcounter{footnote}{0}
\setcounter{page}{1}
\setcounter{section}{0}
\setcounter{subsection}{0}
\setcounter{subsubsection}{0}

%%%%%%%%%%%%%%%%%%%%%%%%%%%%%%%%%%%%%%%%%%%%%%%%%%%%%%%%%%%%%%%%%%%%%%%
\section{Introduction} \label{sec:intro}

Two of the most compelling theoretical constructions since the advent of the
Standard Model (SM) have been the concepts of grand
unification~\cite{unif} and supersymmetry (SUSY).
However it is by now well known that unification, specifically gauge
coupling unification, and supersymmetry
are intimately connected in light of the precise LEP electroweak
data. That the separate gauge couplings of the $SU(3)\times SU(2)\times
U(1)$ SM gauge group unify around the scale $2\times 10^{16}\gev$ if the
SM is embedded in its supersymmetric extension, the minimal supersymmetric
standard model (MSSM), can be taken as the first, albeit indirect, evidence
for SUSY. The initial excitement surrounding this result has been replaced
with a realistic reappraisal of the details of the calculation, showing that
simple unification with a light SUSY spectrum and without modest
grand unified (GUT) scale
corrections predicts $\alpha_3(\mz)$ to be larger than indicated either by
LEP or by low-energy data~\cite{bagger}. Nonetheless there remains a
remarkable level of agreement given the large number of model-dependent
uncertainties that arise in the calculation. 

It is necessary, however, to distinguish which aspects of the MSSM are
fundamental to the observation of gauge coupling unification and which are
coincidental. For example,
consider the addition of extra matter, beyond that of the MSSM,
at some arbitrary scale $\mnew > \mz$. Taking
$\alpha_1=\frac{5}{3}\alpha_Y$ and $\alpha_2$ as measured inputs at
the weak scale, one can use the 1-loop renormalization group equations (RGE's)
for the gauge couplings
to yield a prediction of, \eg, the unification scale, $\mx$:
\beq
\log\left(\frac{\mx}{\mz}\right)=
\frac{2\pi}{b'_2-b'_1}\left(\frac{1}{\alpha_2}-
\frac{1}{\alpha_1}\right)+
\frac{\delta b_2-\delta b_1}{b'_2-b'_1}\log\left(\frac{\mnew}{\mz}\right),
\label{eq:mx}
\eeq
where $b'_i=b^0_i+\delta b_i$ with $b_i^0$ the MSSM $\beta$-function
coefficients and $\delta b_i$ the contributions of the extra matter.
A similar form applies for the modified prediction of $\alpha_3(\mz)$.
Thus we recover the very well known result that, at 1-loop, new states
which shift all three $\beta$-functions identically
(\ie, $\delta b_1=\delta b_2=\delta b_3\equiv\delta b$)
leave unchanged the predictions for the strong coupling and for
the unification scale. Only the value of the experimentally
inaccessible number $\alx$ changes. 

The requirement $\de b_i = \de b$ is met by adding states with
quantum numbers such that they can be thought of as fitting into complete
representations of some simple group containing the SM,
presumably in conjugate pairs to allow vectorlike
$SU(3)\times SU(2)\times U(1)$-preserving mass terms and for
anomaly cancellation. Thus the apparent unification already present
in the MSSM is not simply an accident if there exist only
complete ``GUT'' representations above the weak scale.
{}From the point of view of gauge coupling unification, the MSSM and these
extended variants are on an equivalent footing -- there is no
{\sl current} experiment that favors one over the other.

That said, in this paper we will show that there can exist potentially
dramatic and experimentally observable differences
(at the next generation of colliders) between these models
given access to the sparticle spectrum.
These predictions will provide a channel for detecting this extra matter
through its virtual effects even if it is too heavy to be observed directly.
Specifically, we will study intra-generational sparticle mass ratios and
gaugino mass ratios as discriminants, under the assumption that they
exhibit mass unification at the unification scale typical of
supergravity-mediated models of SUSY breaking.
That masses in the scalar sector are universal is a strong assumption;
however, low-energy flavor-changing neutral current 
constraints require inter-generational universality, so
intra-generational universality seems motivated. Gaugino mass
unification, on the other hand, is very well motivated, both in GUT's and
in string theory, as we will discuss in greater detail below.

Each of these mass ratios possesses
various advantages and disadvantages. The sparticle masses have a quite
sensitive
dependence on the existence of even relatively small amounts of extra matter,
but there are many other contributions to sparticle masses of a fairly
generic nature -- D-terms from broken symmetries, Planck-scale corrections to
the K\"ahler potential, \etc\ -- that may make it difficult to disentangle
the contributions of the extra matter in a unique way. On the other hand,
the gaugino mass ratios, as we will show, are largely immune from corrections
arising from unknown high scale physics.
But we will also see that gaugino mass ratios
differ from their canonical values only slightly
in the presence of small amounts of extra matter; however the ratios
can begin to differ significantly once $\alx>\al_3(\mz)$.

It is an inevitable consequence of adding additional matter to the MSSM
that $\alx$ increases. Specifically, we will concern ourselves
with scenarios of {\sl semi-perturbative unification} (SPU)
in which matter in complete $SU(5)$ multiplets is added at some
intermediate scale $\mnew<\mx$ such that $\alx > \al_3(\mz)$; in order to
trust our results, however, {\sl we require $\alx$ remain perturbative
in the sense of quantum field theory}. A reasonable test of perturbativity
is that the contributions from the $(n+1)$-loop RGE's are small compared to
those at $n$-loops. Because we will be working in a regime in which the
$n=1$-loop contributions can be anomalously small, a test comparing 2-loop
to 1-loop contributions can be misleading. Instead,
we will test for
perturbativity by comparing the 3-loop contributions to those at 2-loops.
This will force $\alx\lsim 1/2$. Note that we do {\sl not} impose
that the gauge coupling remain perturbative all the way up to the
Planck (or reduced Planck) scale, since unification into a string theory
can occur well below that scale, as we will discuss in the next subsection.

Our primary results are two-fold: first we generalize some recent results
of Babu and Pati~\cite{babu}\ to show that SPU pushes up the unification scale,
$\mx$, sometimes significantly, towards our expectation from
string theory. Second, we examine the imprint left on the light SUSY spectrum
by SPU and extra matter in general. We will finally consider threshold
corrections, particularly those at the high scale.

Throughout, we will present our results both
numerically and analytically. Because of the difficulty in finding
simple and general analytic expressions when the 2-loop contributions compete
with
the 1-loop, we will usually confine our analytic results to the interesting
reference case of
$b_3=0$, where the running of $\al_3$ is entirely 2-loop in origin; for
this case (in which $\alx\simeq0.22$) it is obvious that $\alx$ is
perturbative.

This paper is organized as follows: in the rest of this Section we
will discuss SPU scenarios within the context of string theory and the
relationship of SPU to the older idea of non-perturbative unification.
Those allergic to high-scale handwaving are encouraged to jump to
Section~\ref{sec:spu}\ in which we briefly address the question of
the amount and mass scale of the extra matter needed to achieve SPU
of the gauge couplings and discuss the raising of the unification scale
at 2-loops.
Our primary results are contained within Section~\ref{sec:lowE}\ in
which we examine the low-energy consequences of SPU
in the form of novel gaugino mass ratios and squark and slepton spectra.
After our Conclusions we
include a brief Appendix in which the 3-loop RGE's are presented for
the scenarios considered here, based on the recent work of Jack,
\etal~\cite{jj1,jj2}.

\subsection{SPU from a string perspective} \label{sec:mot}

That larger values of the unified coupling $\alx$ may be preferred
can be seen if we view the unification of couplings from the perspective
of string theory. Probably the most serious phenomenological problem
that faces string theory is that of dilaton runaway~\cite{ds}, which
seems to be generic to string theories. In short, to all orders of
perturbation theory (for a supersymmetric string theory) the dilaton,
whose expectation value sets the size of the gauge and other couplings,
has no potential. When ``small'' non-perturbative effects are included (such
as gaugino condensation in a hidden gauge group) a potential
can be generated, but this potential {\sl must vanish} as the dilaton 
vacuum expectation value
goes to infinity, and the theory becomes free. Thus, unless there is
a local minimum at some intermediate value of the dilaton expectation
value, the dilaton either runs away to a free theory, or to a solution
with non-zero cosmological constant, presumably large.
In order to generate a local minimum to stabilize the dilaton
one must almost certainly be outside the region of
``small''  non-perturbative effects, so that one can have competing
terms. If all non-perturbative effects are of field-theoretic
origin (\ie, instanton-like with behaviour $\exp(-8\pi^2/g^2)$), then
this seems to require very large couplings. 

It seems at first to be a disaster that string theory must be 
strongly coupled in order to describe our universe.
Not only does the observed unification within the
MSSM predict a small, perturbative value for $\gstr\sim\gx$,
but the recent results on strong-weak coupling duality suggest that
the dilaton runaway problem just reappears in a new guise if we move
into the strongly-coupled region $\gstr \gg 1$. Thus, from the
duality argument it appears that at best
the value of $\gstr$ is in the region of intermediate coupling
(probably near the electric-magnetic self-dual point $\gstr\sim\sqrt{2\pi}$),
where {\sl field-theoretic} nonperturbative effects are still negligible and
cannot stabilize the dilaton.

A possible solution
to this conundrum may lie in the observation that coupling
strengths which within the context of field theory are perturbatively
small can within the context of string theory be nonperturbatively large.
In string theory there are expected to be corrections,
specifically to the K\"ahler
potential, which grow as fast as $\exp(-a/g)$, where
$a\sim 1$~\cite{shenker}. Thus we might hope that
Nature has chosen $\gstr$ such that $\exp(-a/\gstr)\sim1$ while
$\exp(-8\pi^2/\gstr^2)\ll1$, allowing us to approach unification
in perturbation theory while still understanding the stabilization of
the dilaton. Values of $\gstr$ within the SPU range, $g_3(\mz)<\gstr
\lsim\sqrt{2\pi}$, are certainly within this domain.

In string theory there is also the well-known problem
of the scale of coupling unification. One expects, for string theory,
unification not only among the field theoretic
couplings but also with gravity~\cite{ginsparg}. A 1-loop calculation within
weakly-coupled heterotic string theory yields a prediction for the
scale at which such unification occurs, the string scale,
as a function of the unified coupling, $\gstr$~\cite{kaplunovsky}:
\beq
\mstr=5.3\times\gstr\times 10^{17}\gev,
\eeq
only about one decade from the MSSM unification scale $2\times
10^{16}\gev$ with $\alx\simeq1/25$. But converted
to a prediction for $\al_3(\mz)$ within the MSSM the string result
is many standard deviations away from the experimentally observed
value. There have been many suggested resolutions
to this disagreement~\cite{dfm}, including
the addition of matter in incomplete $SU(5)$ multiplets, the inclusion
of (hopefully large) string-scale threshold corrections, and even non-standard
affine levels for the affine algebras (Kac-Moody algebras) giving
rise to the SM gauge interactions. Most recently the question of the
unification scale has been investigated within the context of
strongly coupled $E_8\times E_8$ string theory~\cite{ed}. The low-energy
limit of this theory is 11-dimensional supergravity with the 11th
dimension being an interval. The length of this interval is essentially a free
parameter which can be fit using $\alx$, $\mx$ and Newton's constant (in
units of the 11-dimensional Planck length, $\ell_{11}$)
so that the unification scales in the string and field theories correspond.
If we take $\alx$ and $\mx$ to be those of the MSSM then the length of
this interval is about $70\ell_{11}$, quite large. This in turn has
potentially interesting consequences for cosmology, axion dynamics,
\etc~\cite{banks}. However larger values of $\alx$ and $\mx$ are in no
way disfavored by this result; they simply lead to different values for
the length of the interval and therefore different phenomenology.

In Section~\ref{sec:spu}, we will show that in SPU the unification
scale is automatically raised by the 2-loop effects, approaching
in some cases the 1-loop prediction of the string scale quite closely.

\subsection{Relation to non-perturbative unification}

Finally, we wish to mention the connections and differences of our
SPU scenarios with an earlier scheme, Non-Perturbative Unification
(NPU), first proposed by Maiani, Parisi and Petronzio~\cite{mpp}\
in 1979. The basic idea of NPU is that
as more and more states are added to the particle spectrum, the
$\beta$-functions for the gauge couplings increase until a Landau
pole at scale $\La$ develops. If unification occurs in the MSSM, then
the amount and mass scale of extra matter can be chosen such that unification
occurs right at $\mx=\Lambda$. The value of $\gx=g(\Lambda)$ then becomes
irrelevant, all low-scale observables depending only on the scale
$\Lambda$ itself. The weak scale values of the gauge couplings
appear as infrared pseudo-fixed points of the renormalization group
equations (RGE's).

A large number of  analyses have been performed of the NPU scheme in the
extended MSSM, most very similar in nature. These analyses
have three rather generic problems: {\sl (i)} they assume that the gauge
couplings become nonperturbative at the unification scale which is no more
pleasant for string theory than very weak couplings, {\sl (ii)}
their only tool for analysing the unification is perturbative RGE's, used
despite the fact that the unification is supposedly non-perturbative. And
{\sl (iii)} thanks to the non-perturbative nature of the couplings,
other observables such as scalar and gaugino masses are assumed
to have uncontrollable corrections at the unification scale which prevent
any prediction of their values at the weak scale. Thus the only
discriminating signal of NPU is to actually find the extra matter through
on-shell production. Within SPU, we will see
that the coupling strengths necessary in order to render interesting
effects at the
weak scale observable are of intermediate strength. Moreover we will have
control of the perturbative expansion by checking against the
3-loop contributions.

\section{Semi-perturbative unification} \label{sec:spu}

In this work, we assume that nature chooses
to unify semi-perturbatively. Therefore the low-energy values of the
gauge couplings which are measured experimentally are by definition
close to their infrared pseudo-fixed point values and have their
measured values thanks to some combination of extra
matter at some unknown scales
and possibly new large Yukawa couplings involving that extra matter.
We have no knowledge {\sl a priori} of these dynamics but hope to study
those effects which are independent of the details. Therefore we will
be interested in increasing the $\beta$-functions until the unification
scale is pushed close to, but not above, the Landau scale and examining
the resulting phenomenology.

All of our methods for analysing physics within this domain will be
perturbative. Each result derived perturbatively must then be checked
against some test of perturbativity to ensure its validity. As already
mentioned, a good test for a result derived at $n$-loops would be a
calculation of the $(n+1)$-loop corrections. This test does not work for
$n=1$ for two reasons: the 1-loop $\beta$-function for $\al_3$ is
anomalously small in the region of interest, and many of the effects in which
we are interested only arise at 2-loops. Therefore we will use as our test
of perturbativity the ratio $|\beta_i^{(3)}/\beta_i^{(2)}|$ for each gauge
group $i$, where $\beta^{(n)}$ is the $n$-loop gauge $\beta$-function.
For the purposes of this study, in calculating $\beta^{(3)}$ we will set all
Yukawa couplings to zero; see the Appendix for a full discussion of the
relevant RGE's.
We will make the somewhat arbitrary, but reasonable, choice that the
perturbative expansion is valid if
$|\beta_i^{(3)}(\mx)/\beta_i^{(2)}(\mx)|\lsim1/2$ for
all $i$. All of our results are derived under this constraint.

We also, of course, assume that the near-unification of the three
gauge couplings in the MSSM is not an accident. For the purposes
of our calculations we will denote as $\mx$ the scale at
which $\al_1(Q)=\al_2(Q)$ and determine the corresponding value of
$\al_3(\mz)$ as a prediction by running back down to the weak
scale. Since we only allow complete GUT multiplets to
be added to the MSSM, we know that we cannot disrupt the full unification
that occurs in the MSSM by much.\footnote{The
precise value of $\al_3(\mz)$ is not a particularly useful prediction of
SPU (or the MSSM for that matter) without considering the corrections at
the weak scale, logarithmic and non-logarithmic, which are known to
be large~\protect\cite{bagger}. In this sense, we are not requiring
precise unification of $\al_3$ with $\al_1$ and $\al_2$.
Furthermore, note that shifts in $\al_3(\mz)$ which arise due to
the extra matter are typically cancelled against those induced by
splittings in the masses of the new matter generated by their
anomalous dimensions~\protect\cite{carone}.}
We also note in passing that the
other canonical prediction of minimal GUT's, $b$-$\tau$ Yukawa unification,
works in the MSSM only as one of the fermion Yukawas approaches its
infrared pseudo-fixed point, and does not work at all once extra
matter is included; we will not consider $b$-$\tau$ unification further.

As we do not have control over the specifics of the dynamics which are
occurring between the weak and unification scales, we need an appropriate
parametrization for describing the unknown effects.
If the effective theory at the weak scale is the MSSM then there are
essentially only two degrees of freedom for exploring SPU: the representations
of the extra matter and the mass scale at which they couple. Consider the
toy case where the new matter is degenerate at the weak scale, $\mnew=\mz$.
We can derive
bounds on $\delta b$ by requiring that $1/\alx$ approach zero from above.
Thus at 1-loop:
\beq
\delta b\lsim \left(b^0_1-\frac{\al_2}{\al_1}b^0_2\right)
\left(\frac{\al_2}{\al_1}-1\right)^{-1}\simeq4.6.
\label{eq:maxb}
\eeq
A $\five$ and $\ten$ of $SU(5)$ contribute 1/2 and 3/2 to $\delta b$
respectively, while a $\sixteen$ of $SO(10)$ contributes 2.
Eq.~(\ref{eq:maxb}) then sets the maximum number of additional
$\five$'s, $\ten$'s and $\sixteen$'s at 9, 3 and 2 respectively.
There are no other representations which can be added at the weak scale.
If $\mnew\gg\mz$ then there can be correspondingly more states added, or
alternatively larger GUT representations.

It seems then that an effective
number of $\five$'s, $\ten$'s or $\sixteen$'s
added to the model at the weak scale may be a good parametrization for
studying the new effects of SPU. In particular, we will study how the
phenomenology changes as the effective number of representations is increased
to the SPU point. 
We will choose the {\sl effective} number of $\five$'s, $\ten$'s or
$\sixteen$'s as the degree of
freedom in most cases, and differentiate it from the {\sl actual}
number, $n_i$, by denoting it $\nieff$, for $i=5$, 10 or 16.
Since we are absorbing not only the number of extra
representations into $\nieff$ but also their mass scale and other
effects (see below),
{\sl it is not necessary that $\nieff$ be an integer}.

\begin{figure}
\centering
\epsfysize=3in
\hspace*{0in}
\epsffile{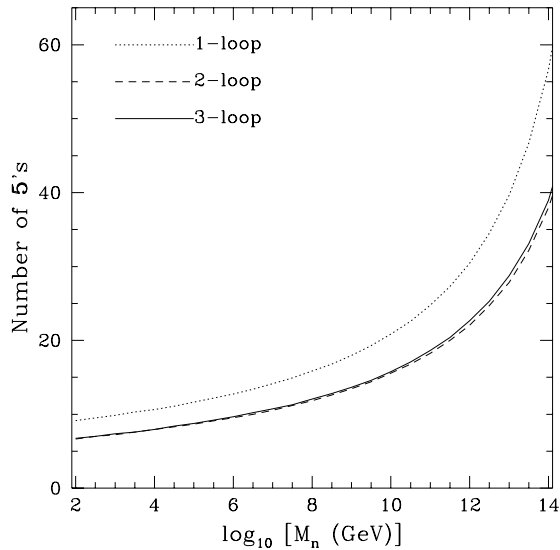}
\caption{Dependence of the maximal 
number of $\five$'s on their mass scale, calculated
at 1-, 2-, and 3-loops, and with all Yukawas set to zero.}
\label{fig:nmax}
\end{figure}

There are two dominant effects which change the number of representations
that can/must be added to the MSSM at a given scale $\mnew$ to achieve SPU.
Two-loop contributions to the RGE's tend to increase the gauge
$\beta$-functions. In the MSSM (with or without extra matter) this
decreases the amount of matter that can/must be added compared to the
1-loop case.
In Figure~\ref{fig:nmax}\ we plot the extreme upper bound on $\nf$ as
a function of $\mnew$. (By ``extreme upper bound'' we mean the value beyond
which
the Landau scale occurs below the unification scale, calculated to the stated
order in perturbation theory. This is not to be confused with our usual
definition of upper bound which requires $|\beta_i^{(3)}/\beta_i^{(2)}|<1/2$
at the unification scale.) Results are shown at 1-, 2- and 3-loops,
where for the puposes of the figure all Yukawa couplings are set to zero.
In the 1-loop case one can rescale the $y$-axis to $\nt$ or
$n_{16}$ using the relation $\nf=3\nt=4n_{16}$; the 2-loop corrections do
not have any such simple scaling. Note that $\delta n\simeq 3$ or 4 
in going from
1-loop to 2-loops over most of the range of $\mnew$.
%%%%%%%%%%%%%%%%%%%%%%%%
%%%%%%%%%%%%%%%%%%%%%%%%

Yukawa couplings enter the 2-loop $\beta$-functions with
opposite signs from the gauge contributions and therefore slow the running
of the gauge couplings, an effect
which might either increase or decrease the amount of extra matter which
is allowed. One can parametrize this effect as a shift in $\nieff$
away from $n_i$; that is, the
effects of the Yukawas mimic extra matter:
\beq
\beta_i=\alpha_i^2\left(\frac{b_i}{2\pi}-\sum_j\frac{a_{ib}}{32\pi^3}
y_b^2\right)+\CO(\alpha^3)
\eeq
for some Yukawa coupling $y_b$. Ignoring the running of the Yukawas themselves
and generalizing Eq.~(\ref{eq:maxb}), one finds\footnote{For each observable 
the precise definition of $\nieff$ in terms of $n_i$,
the new mass scale, and any other effects such as Yukawas differs slightly. 
This particular definition is appropriate for parametrizing effects of Yukawas
on $\alx$ and thus on the amount of extra matter needed for SPU.}:
\beq
\delta\nfeff=\sum_b\frac{y_b^2}{8\pi^2}\left(a_{1b}-\frac{\alpha_2}{\alpha_1}
a_{2b}\right)\left(\frac{\al_2}{\al_1}-1\right)^{-1}
\label{eq:neff}
\eeq
For Yukawas like those of the SM particles (including the top Yukawa),
$\nfeff<0$ so more matter can be added to the spectrum and still
maintain perturbativity. This is not true for generic Yukawa couplings;
in particular, most $R$-parity violating couplings lead to $\nfeff>0$,
though their coefficients are usually assumed very small.

Although SPU has no effect on the scale of unification at 1-loop,
at 2-loops this no longer holds. For small values of $\alx$ the 2-loop
effects are calculable and negligible, but for SPU, the shifts in the
unification scale can be substantial. Consider, for example, our reference
case of $b_3=0$. It is necessary to solve the $\al_3$ equation
\beq
{d\al_3\over dt} =
\frac{6}{\pi^2}\al_3^3 + \cdots
\label{eq:al3form}
\eeq
where the ellipses represent subleading terms which go as
$\al_3^2\al_{1,2}$ and are also down by small coefficients.
Making the good approximation of dropping these
terms, we get the solution
\beq
\frac{1}{\al_3^2(\mu)} \simeq \frac{1}{\al_3^2(\mz)} - \frac{12}{\pi^2} \log
\left(\frac{\mu}{\mz}\right),
\label{eq:al3sol}
\eeq
so $\alx>\alsz$ as expected. With $\alsz=0.12$ this leads to
$\alx\simeq 0.22$. In this case the expression for the
unification scale becomes
\beq
\mx\simeq\mx^{(1)}\left[\prod_{i=1,2}\left(\frac{\alx}{\alpha_i}\right)^{\frac
{b_{2i}-b_{1i}}{2b_i(b_1-b_2)}}\right]
\exp\left(\frac{\pi}{24}\frac{b_{23}-b_{13}}{b_1-b_2}\left(\frac{1}{\alsz} -
\frac{1}{\alx}\right)
\right),
\label{eq:mx3}
\eeq
where $\mx^{(1)}$ is the 1-loop unification scale, about
$(2\sim3)\times10^{16}\gev$.
Note that for SPU $(1/\alsz - 1/\alx)>0$ and $b_1-b_2 = 28/5$,
but the $b_{ij}$ depend on the type of matter added to set $b_3=0$, \ie,
either 2 $\ten$'s or 6 $\five$'s. For either case the first factor
coming from the $U(1)$ and $SU(2)$ contributions raises the unification scale
by a factor $\sim4$. On the other hand, the exponential
($SU(3)$ contribution) depends strongly on type of additional matter
since,
\beq
b_{23}-b_{13} = \biggl\{ \begin{array}{ll}
64/5 & \mbox{if $\nt=2$} \\
0 & \mbox{if $\nf=6$,}
\end{array}
\label{eq:b2b1val}
\eeq
which results in an additional enhancement
in Eq.~(\ref{eq:mx3}) of $\sim 3$ in the case of $\ten$'s, but none
in the case of the $\five$'s. This is due to the presence of $(\bf{3,2})$
states in the decomposition of the $\bf{10}$ which lead to enhanced
$b_{23}$ entries in the 2-loop $\be$-function coefficients.
Thus there is a quantitative difference at 2-loops between adding
(1-loop) equivalent amounts of $\five$'s and $\ten$'s.

\begin{figure}
\centering
\epsfxsize=5.75in
\hspace*{0in}
\epsffile{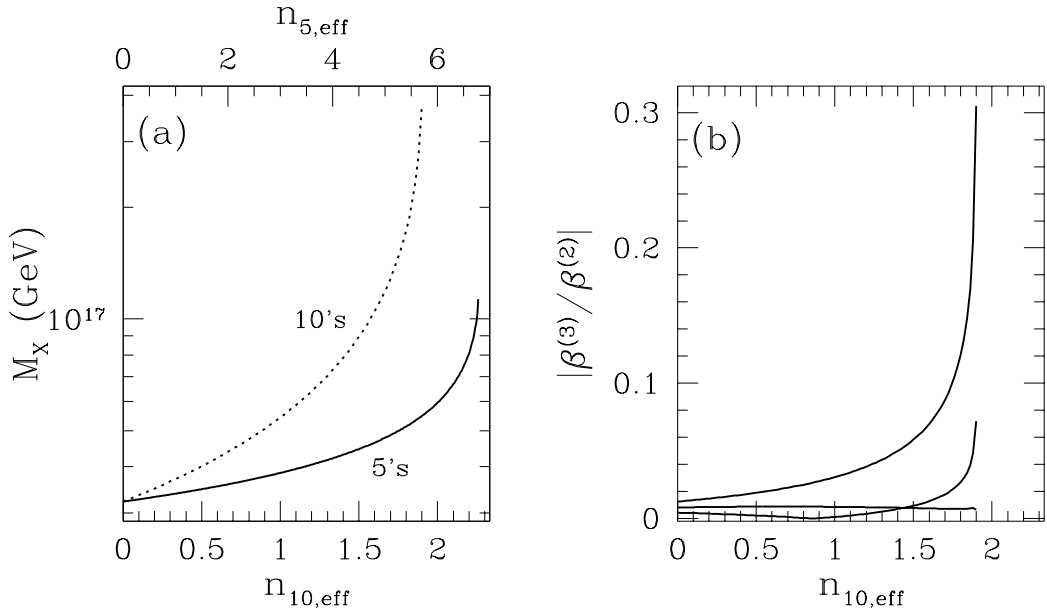}
\caption{(a) Dependence of the unification scale on the type and amount of
extra matter; (b) ratio of the 3-loop to 2-loop contributions to the
three $\beta$-functions at the unification scale with extra $\ten$'s. 
The ratio for the $U(1)$
coupling provides the strongest constraint, followed by that of the $SU(2)$.}
\label{fig:unifplot}
\end{figure}

In Figure~\ref{fig:unifplot}(a) we show a full 3-loop numerical
calculation of the unification scale as function of
$\nfeff$ and $\nteff$. This clearly shows the increase in the unification
scale for both $\five$'s and $\ten$'s, and that the increase is more
marked in the second case. In line with the analytic estimates in
Eq.~(\ref{eq:mx3}), the unification scale for $\ten$'s is about a factor
of 3 higher than that for $\five$'s. It is quite remarkable that the
unification scale in these models, especially in the case of $\ten$'s,
approaches quite closely the 1-loop heterotic string prediction
($\sim 1\times10^{18}\gev$ for the appropriate value of $\gstr$).
Note that this occurs without the introduction of split multiplets or large
weak or string scale threshold corrections. 

In Figure~\ref{fig:unifplot}(b) we plot the ratio
of the 3-loop term to the 2-loop term evaluated at the unification
scale for the three SM gauge couplings. Notice that the perturbative expansion
breaks down first in $\beta_1$, with $|\beta_1^{(3)}/\beta_1^{(2)}|$
reaching values near/below 1/2 as we approach the cutoff in the amount of extra
matter. We take this as a strong indication 
that our perturbative calculations are under control.
%%%%%%%%%%%%%%%%%%%%%%%%
%%%%%%%%%%%%%%%%%%%%%%%%

As the scale of extra matter increases, more matter is needed to reach SPU.
However the prediction of $\mx$ remains roughly constant, as we have checked
numerically. We also note the the case of the $\sixteen$'s is intermediate
to those of the $\five$'s and $\ten$'s, as one might expect.

\section{Low-energy signatures} \label{sec:lowE}

The low-energy signatures of SPU on which we focus all involve
changes to the spectra of sparticle masses at the weak scale.
Further, all the statements that we make in this regard will
be in the context of supergravity (SUGRA) mediated
supersymmetry breaking scenarios~\cite{kkrw}. The reason for this
restriction is that the main effect of the additional
matter on sparticle masses will be radiative, through the modified
RGE running of the soft SUSY breaking parameters, and this
only occurs if the soft masses are induced above the scale
of the additional matter. Therefore we will, for example, have
nothing to say about the case of gauge-mediated SUSY breaking
where the scale at which the soft terms are induced in the observable
sector is very close to the scale of the additional messenger
matter.

In line with the usual assumptions, we will take the soft
terms induced at the unification scale to be {\sl universal}
in form. This is certainly a strong assumption for the soft
scalar masses, but one of our main points will be
that the usual low-energy predictions of such a scenario
can be greatly altered even without violations of
universalty at the high scale. Futhermore, for the {\sl gaugino}
masses the universality assumption is relatively mild,
as we will review below.

Probably the single most interesting and distinctive signature
of SPU, at least near the upper limit of the allowed range of
unified coupling, is the change in the low-energy gaugino
mass ratios. Recall the usual situation within
SUGRA-mediated SUSY breaking, where at the
unification scale we expect universal gaugino masses
\beq
M_1 =M_2=M_3 = \mhalf.
\label{eq:univgau}
\eeq
Given this boundary condition the low-energy ratios are determined
by the running from $\mx$ down to the weak scale. The
2-loop RGE's for the gaugino masses are very close in form to those of the
gauge couplings:
\beq
{d M_i\over dt} = \frac{b_i}{2\pi}\al_i M_i +
\frac{b_{ij}}{8\pi^2}\al_i\al_j(M_i + M_j) +
\frac{a_{ic}}{32\pi^3}\al_i y_c^2 M_i +\cdots,
\label{eq:gauginorg}
\eeq
where the ellipses represent $A$-term contributions which can be
shown to be small both in the MSSM and with extra matter (see the work of
Yamada in Ref.~\cite{gauginorge}).
The $\be$-function coefficients $b_i$, $b_{ij}$ and $a_{ic}$ are
equal to those for the gauge couplings given in the
Appendix~\cite{gauginorge}.
Such a form for the gaugino RGE's implies that the ratio
\beq
R_i\equiv M_i/\al_i
\label{eq:Rratio}
\eeq
satisfies the equation
\beq
{dR_i\over dt} = \frac{b_{ij}}{8\pi^2}\al_j^2 R_j + \cdots
\label{eq:Rrgeform}
\eeq
where again the ellipses represent the small $A$-term contributions.
In other words the ratio is constant at 1-loop but runs at 2-loops.
In the case of the unextended MSSM the change in the ratio
due to the 2-loop term is quite small, and we get the standard
result that at the weak scale
\beq
M_i/M_j = \al_i/\al_j
\label{eq:usualR}
\eeq
up to relatively small weak-scale threshold corrections and
conversions from $\ol{{\rm DR}}$ masses to pole masses.
Thus $M_3:M_2:M_1 \sim 7.4:2.0:1.0$.

In the SPU scenario these ratios can change dramatically.
In our reference case with $b_3=0$, the equation
for $R_3$ has an $\al_3^2R_3$ term with large coefficient
and $\al_i^2R_i$ terms ($i=1,2$) that are suppressed both by
relatively small coefficients and the fact that below $\mx$ the
$U(1)$ and $SU(2)$ couplings decrease quite quickly.
Keeping just the dominant $\al_3^2R_3$ term leads to the
expression (valid to roughly 10\%)
\beq
R_3(\mz) \simeq \frac{\mhalf}{\alx}\left(\frac{\alsz}{\alx}\right).
\label{eq:R3pred}
\eeq
The term in parentheses is the modification to the usual result
and amounts to a 40\% change in the predicted
value of $R_3$ as compared to the MSSM.

For the other ratios, $R_{1,2}$, a similar approximation scheme
is applicable. The $\al_3^2R_3$ term again dominates for most of
the running, although the $\al_{1,2}^2R_{1,2}$ terms can provide a
numerically significant
correction near the unification scale. These can be simply dealt
with by substituting in the 1-loop expressions and integrating.
The general form of the prediction for the ratios $R_1$ and $R_2$
is then
\beq
R_i(\mz) \simeq \frac{\mhalf}{\alx}\left\{ 1 - B_i - \sum_{j=1,2}
\frac{b_{ij}}{4\pi b_j}\alx \right\} + B_i R_3(\mz),
\label{eq:Rjpred}
\eeq
where the constants $B_i=b_{i3}/b_{33}$ are ratios of two 2-loop
$\be$-function coefficients that depend upon the type of extra matter.
We have also dropped small correction terms of order
$\mhalf\al_i(\mz)/4\pi\alx$. The form of Eq.~(\ref{eq:Rjpred}) is actually
{\sl valid beyond the particular case of} $b_3=0$ -- it is the
generalization of the usual one loop relation $R_i=\mhalf/\alx$
to 2-loops.

For $R_2$ in the specific case of $b_3=0$ this leads to the prediction
\beq
R_2(\mz) \simeq \frac{\mhalf}{\alx}\left\{ 1 - B_2 - A_2\alx \right\}
+ B_2 R_3(\mz),
\label{eq:R2pred}
\eeq
where the constants $A_2$ and $B_2$ take on
the values $281/96\pi$ and $5/6$ ($95/32\pi$ and $1/2$) in the
$\ten$ ($\five$) cases respectively.

The ratio $R_1$ may be handled in an identical way leading to
the relation
\beq
R_1(\mz) = \frac{\mhalf}{\alx}\left\{ 1 - B_1 - A_1\alx \right\}
+ B_1 R_3(\mz),
\label{eq:R1pred}
\eeq
in the $b_3=0$ case, with $A_1= 337/480\pi$ $(147/160\pi)$ and
$B_1=b_{13}/b_{33}= 17/30$ $(1/2)$, in the $\ten$ ($\five$) cases.

\begin{figure}
\centering
\epsfxsize=5.75in
\hspace*{0in}
\epsffile{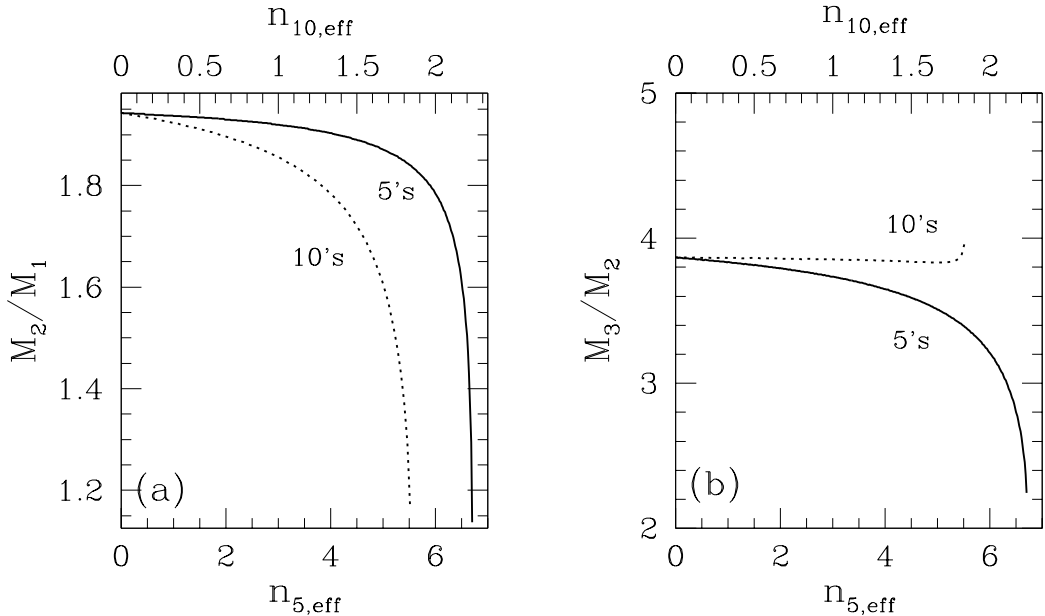}
\caption{Ratios of the gaugino masses: (a) $M_2/M_1$ and (b) $M_3/M_2$ as
functions of the amount and type of extra matter. Solid lines are for the case
of additional $\five$'s, dotted lines for $\ten$'s.}
\label{fig:gauginos}
\end{figure}

In Figure~\ref{fig:gauginos}
we show the results of the numerical evaluation of
tha low-energy gaugino mass ratios (a) $M_2/M_1$ and (b) $M_3/M_2$ as
a function of $\nfeff$ and $\nteff$.
The most remarkable feature of the Figure is that with SPU one can have
large changes in the gaugino mass ratios away from the canonical values
of $\al_i/\al_j$. In particular, $M_2/M_1$ can approach unity,
depending on the type of extra matter. This has strong phenomenological
consequences. One of the neutralinos of the MSSM can be essentially
photino-like, rather than bino-like as is usually assumed. If that neutralino
is the lightest SUSY particle then it can be the dark matter in the universe
with properties markedly different than one might expect from bino-like
dark matter~\cite{cooper}. 
For example, because it lacks a coupling to the Higgs it does not 
self-annihilate as efficiently, resulting in a higher relic density than for
a similarly massive bino.

As a way to differentiate the cases of the $\five$'s and $\ten$'s it would be
useful to have access to the $M_3/M_2$ mass ratio. Figure~\ref{fig:gauginos}(b)
reveals that as one approaches SPU, $M_3/M_2$ remains constant or slightly
increases 
for $\ten$'s while it decreases for $\five$'s; the behavior for the $\five$'s
is well described by Eqs.~(\ref{eq:R3pred})--(\ref{eq:R1pred}), but that
of the $ten$'s requires a much more detailed analytic analysis because of
cancellations among competing terms. 
(Note that Figure~\ref{fig:gauginos}
was made assuming that $\al3(\mz)$ is brought back to the experimentally
measured values using threshold effects.)
%%%%%%%%%%%%%%%%%%%%%%%%
%%%%%%%%%%%%%%%%%%%%%%%%

One may worry that these predictions for the gaugino mass
ratios suffer from large uncertainties due to threshold
corrections, either at the low, or especially at the high scale,
due to the large amount of matter present. However this is not
the case. Part of the reason is obvious: because the 1-loop
running of the gaugino masses and the gauge couplings are
identical, all logarithmically-enhanced threshold terms
such as
\beq
\frac{\alx}{4\pi}\log\left(\frac{M_V^2}{\mx^2}\right)\quad\mbox{or}\quad
\frac{\alx}{4\pi}\log\left(\frac{M_c^2}{\mx^2}\right),
\label{eq:thresterms}
\eeq
where $M_V$ ($M_c$) is the mass of some superheavy vector (chiral)
multiplet, cancel in the ratio $M_i/\al_i$. Actually the situation
is even better than this. The 1-loop non-logarithmically-enhanced
threshold corrections to the gaugino masses have been calculated in
Ref.~\cite{hisano}, resulting in the expression
\beq
\frac{M_i(\mu)}{\al_i(\mu)} =
\frac{M_{1/2}}{\alx} + \frac{1}{4\pi}\left\{ 2T_i^{(V)}(M_{1/2}(\mu)
-\de m) + \sum_{c={\rm chiral}} T_i(R_c) B_c \right\},
\label{eq:nonenhanced}
\eeq
where $\de m$ is the mass of the fermion component of the Nambu-Goldstone
multiplet induced by SUSY breaking (and is $\CO(\mz)$),
the $B_c$ are the standard
$B$-terms for the chiral multiplets, and the $T_i$ group factors are
defined in the Appendix.
In the case of universal scalar mass terms (and $B$ parameters)
the contribution of complete GUT multiplets to $T_i(R_c) B_c$
add equally to each ratio $M_i/\al_i$, and are
thus harmless in the $M_i/M_j$ ratios. The only non-vanishing contribution
from chiral multiplets in the universal case arises from heavy Higgs triplets,
and is thus independent of the amount of extra matter.
Similarly there is a small
correction arising from the heavy {\sl vector} multiplets, whose
contribution only depends on the gauge structure of the underlying
theory. The final result is a total threshold correction to the gaugino mass
ratios of only a few percent, independent of any unsplit chiral
multiplets at the high scale. Thus high scale field-theoretic
corrections to our expressions are generically under control.

Gaugino mass unification is also a generic prediction of string theory.
One-loop perturbative string threshold corrections to universality have
been considered and argued to be small except in the limit where the
moduli $F$-terms are much larger than those of the dilaton.
There is also the question of non-perturbative corrections to gaugino
masses in string theory. Banks and Dine~\cite{banksdine}
have argued that through a combination
of the holomorphy of the gauge kinetic functions, $f_i$, and discrete gauged
subgroups of Peccei-Quinn symmetries one can show that string non-perturbative
corrections to gaugino masses behave as $\exp(-8\pi^2/g^2)$, of the same
order as field theoretic non-perturbative effects which we are by the
definition of SPU taken to be small. Therefore string-induced
corrections to our expressions are also generically under control.

We now turn to the other major low-energy signal for a larger
value of the unified coupling, the squark and slepton spectrum.
One interesting feature of the modified spectrum is that
as a function of the amount of additional matter, the shifts
in the squark and slepton masses occur well before the maximal SPU point
is reached.  It is therefore sufficient to consider only the 1-loop
equations for the running of these parameters to gain a good
undestanding of the changes. In our analytic results we will also
concentrate on the first two generations of squarks and sleptons
so as to avoid complications due to the large top Yukawa; however our
numerical results include all MSSM Yukawa contributions.
The general form for the 1-loop squark and slepton RGE's is
\beq
{d m_c^2\over dt} = -\frac{1}{2\pi}\sum_i\ga^{(c)}_i\al_i M_i^2,
\label{eq:softrgeform}
\eeq
where the anomalous dimension coefficients $\ga^{(c)}_i$ are not
modified by the addition of extra matter.  They take the values
$\ga^{(Q)} = (1/15,3,16/3)$ for the $\tilde Q$ squark doublets,
$\ga^{(u)} = (16/15,0,16/3)$ for the $\tilde u_R$ squarks,
$\ga^{(d)} = (4/15,0,16/3)$ for the $\tilde d_R$ squarks,
$\ga^{(L)} = (3/5,3,0)$ for the $\tilde L$ slepton doublets, and
$\ga^{(e)} = (12/5,0,0)$ for the $\tilde e_R$ sleptons.
The only dependence
on $\nf$ and $\nt$ comes through the running of $\al_i$ and
$M_i^2$. In the case of the MSSM we can solve these RGE's
by using the 1-loop relation $M_i/\al_i = \mhalf/\alx$, leading
to
\beq
m_c^2(\mz) = m_0^2 + \sum_{i=1}^{3} \frac{\ga^{(c)}_i}{2b_i} \mhalf^2
\left\{ 1 - \frac{\al_i^2(\mz)}{\alx^2} \right\},
\label{eq:sfermionsol}
\eeq
where we have assumed $b_i\neq 0$, and $m_0$ is the soft
SUSY-breaking scalar mass communicated by supergravity at the
high scale (assumed universal for simplicity). Explicitly
\bea
m^2_{\tilde Q} & = & m_0^2 + \mhalf^2\left\{ -\frac{8}{9}(1-\al_3^2/\alx^2)
+ \frac{3}{2}(1-\al_2^2/\alx^2) + \frac{1}{198}(1-\al_1^2/\alx^2)\right\} \\
m^2_{\tilde u} & = & m_0^2 + \mhalf^2\left\{ -\frac{8}{9}(1-\al_3^2/\alx^2)
+ \frac{8}{99}(1-\al_1^2/\alx^2)\right\} \\
m^2_{\tilde d} & = & m_0^2 + \mhalf^2\left\{ -\frac{8}{9}(1-\al_3^2/\alx^2)
+ \frac{2}{99}(1-\al_1^2/\alx^2)\right\},
\label{eq:squarks}
\eea
for the first two generation squarks, and
\bea
m^2_{\tilde L} & = & m_0^2 + \mhalf^2\left\{ \frac{3}{2}(1-\al_2^2/\alx^2)
+ \frac{1}{22}(1-\al_1^2/\alx^2)\right\} \\
m^2_{\tilde e} & = & m_0^2 + \mhalf^2\left\{ \frac{2}{11}(1-\al_1^2/\alx^2)
\right\},
\label{eq:sleptons}
\eea
for the sleptons. An important
qualitative feature of these solutions in the MSSM
is that the $SU(3)$ terms dominate because
of the large ratio $\al_i^2(\mz)/\alx^2\sim 9.8$. This
enhancement of the $SU(3)$ contributions relative to those
arising from $SU(2)$ and $U(1)$ is due to the fact that with
the standard MSSM spectrum the color coupling is still
asymptotically free, while the $SU(2)$ and $U(1)$ couplings
are not. Therefore as we approach the point where $\al_3$ is
no longer asymptotically free, we expect a
substantial compression of the squark and slepton spectrum.
Specifically, the form of the contribution due the $SU(3)$
quantum numbers of the states for our reference case ($b_3=0$)
is modified to
\beq
\De_3 m_c^2(\mz) = \frac{\ga^{(c)}_3\pi}{36} \frac{\mhalf^2}{\alx}
\left\{ 1 - \frac{\al_3^3(\mz)}{\alx^3} \right\}.
\label{eq:modsu3}
\eeq
This leads to a prediction for the masses in the $b_3=0$ case
(left column) as compared to the MSSM (right column) of
\beq
\begin{array}{lc|cl}
m^2_{\tilde Q}  =  m_0^2 + (2.1)\,\mhalf^2 & \quad & \quad &
m^2_{\tilde Q}  =  m_0^2 + (7.2)\,\mhalf^2 \\
m^2_{\tilde u}  =  m_0^2 + (1.8)\,\mhalf^2 & & &
m^2_{\tilde u}  =  m_0^2 + (6.7)\,\mhalf^2 \\
m^2_{\tilde d}  =  m_0^2 + (1.75)\,\mhalf^2 & & &
m^2_{\tilde d}  =  m_0^2 + (6.7)\,\mhalf^2 \\
m^2_{\tilde L}  =  m_0^2 + (0.4)\,\mhalf^2 & & &
m^2_{\tilde L}  =  m_0^2 + (0.5)\,\mhalf^2 \\
m^2_{\tilde e}  =  m_0^2 + (0.12)\,\mhalf^2 & & &
m^2_{\tilde e}  =  m_0^2 + (0.15)\,\mhalf^2
\end{array}
\label{eq:modsfermions}
\eeq
This compression of the squarks down towards the sleptons is
a general signature of SPU that sets in well before the
non-perturbative limit at the high scale is reached, and does
not depend on large 2-loop contributions.

However, to assess the overall scale of the squark and slepton spectrum
it is necessary to re-express $\mhalf$ in terms of the physically
observable gaugino masses. In particular from Eq.~(\ref{eq:R3pred})
we find that (in the case $b_3=0$) $\mhalf\simeq 2.9 M_3$ (versus
$\mhalf\simeq0.33 M_3$ in the MSSM). Written in terms of the gluino mass
parameter, Eq.~(\ref{eq:modsfermions}) becomes:
\beq
\begin{array}{lc|cl}
m^2_{\tilde Q}  =  m_0^2 + (17.6)\,M_3^2 & \quad & \quad &
m^2_{\tilde Q}  =  m_0^2 + (2.4)\,M_3^2 \\
m^2_{\tilde u}  =  m_0^2 + (15.1)\,M_3^2 & & &
m^2_{\tilde u}  =  m_0^2 + (2.2)\,M_3^2 \\
m^2_{\tilde d}  =  m_0^2 + (14.7)\,M_3^2 & & &
m^2_{\tilde d}  =  m_0^2 + (2.2)\,M_3^2 \\
m^2_{\tilde L}  =  m_0^2 + (3.4)\,M_3^2 & & &
m^2_{\tilde L}  =  m_0^2 + (0.17)\,M_3^2 \\
m^2_{\tilde e}  =  m_0^2 + (1.0)\,M_3^2 & & &
m^2_{\tilde e}  =  m_0^2 + (0.05)\,M_3^2
\end{array}
\label{eq:modsfermions2}
\eeq
Given the
bound on the gluino mass of roughly $180\gev$ from CDF~\cite{cdfgluino}\ (this
is in the limit of heavy squarks which is appropriate here),
we find that first two
generations of squark doublets have a mass of at least $750\gev$ in our
reference case,
while the sleptons range in mass from $180\gev$ to $330\gev$, always assuming
$m_0=0$. (The right-handed top squark, because of the large top quark Yukawa,
has a reduced mass relative to the other squarks -- numerically we find its
lower bound to be $500\gev$, ignoring left-right mixing.)
Therefore the squark and slepton spectrum has to be heavier than
is apparent in Eq.~(\ref{eq:modsfermions}).

\begin{figure}
\centering
\epsfysize=3in
\hspace*{0in}
\epsffile{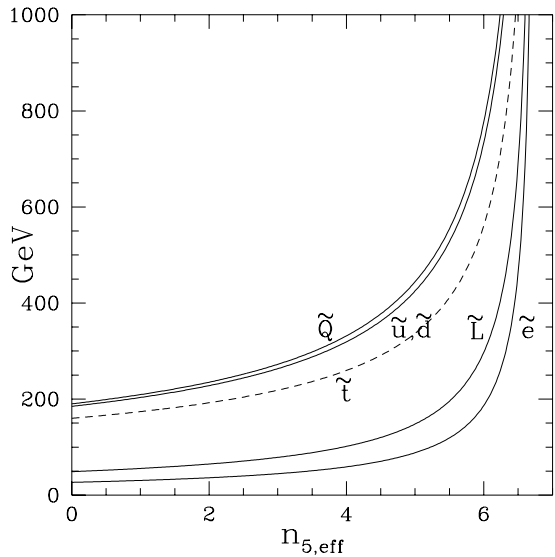}
\caption{Squark and slepton masses as a function of the number of additional
$\five$'s, for constant $M_3=200\gev$ and $m_0=0$. The $\tilde{t}$ is shown
as a dashed line. The $\tilde{u}$ and $\tilde{d}$ contours are coincident.}
\label{fig:sparticles}
\end{figure}

The essential physics is demonstrated in Figure~\ref{fig:sparticles},
where we have taken a constant value of $M_3=200\gev$ and $\tan\beta=2$
and shown how the
scalar masses change as a function of the amount of extra $\five$'s.
(The case of extra $\ten$'s is essentially identical.)
In the Figure are plotted the first/second-generation squarks and sleptons
and the right-handed top squark, $\tilde{t}$, which falls significantly
below the other squarks due to the large top Yukawa. The Figure clearly shows
the overall lifting of the scalar masses with respect to the gauginos; the
compression of the scalar mass ratios is also present but more difficult
to see.
%%%%%%%%%%%%%%%%%%%%%%%%
%%%%%%%%%%%%%%%%%%%%%%%%

It is well-known that in the MSSM electro-weak symmetry breaking (EWSB)
is induced when the $({\rm mass})^2$ of one of the Higgs doublets is
driven negative by radiative effects enhanced by the large top quark Yukawa.
We have studied this question numerically and found that this physics
is qualitatively unaffected by the inclusion of extra matter. 
In particular, the $|m_{H_u}^2|$ scales with $\nfeff$ in a similar fashion 
as the other scalars. Therefore we expect the value of $\mu$ in the MSSM
superpotential to be much larger than the gaugino masses, so that the lightest
SUSY state in this SPU scenario will be a neutralino which is dominantly
photino-like.

\section{Conclusions} \label{sec:concl}

In this paper we have considered the possibility that the gauge couplings
unify within the semi-perturbative regime at high scales. Although
such scenarios are from an experimental viewpoint
currently on an equivalent footing to the MSSM, we showed that they can lead
to striking experimental signatures. In contrast to previous studies
of non-perturbative unification we have been able to make reliable
predictions in our scenario by utilizing the 3-loop 
gauge coupling RGE's as a test of the sensitivity of our predictions to 
higher loop effects.

The addition of extra matter
changes the usual spectrum of scalar masses which one derives from
minimal supergravity-mediated models of SUSY breaking, and more surprisingly,
shifts significantly the relations among the gaugino masses at the weak scale.
In particular, we find that $M_2/M_1\simeq1$ can be achieved for values
of the unified coupling for which the field theory is still perturbative.
This can lead to a host of phenomologically interesting effects coming from
the photino-like nature of the lightest SUSY state. Interestingly, it may
be possible to use
some observables (\eg, $M_3/M_2$) as discriminants among the various
types of extra matter.

We have also demonstrated that a generic prediction of SPU is the raising of
the unification scale well above the canonical value of the MSSM. For
$\ten$'s in particular we find $\mx\approx4\times10^{17}\gev$, remarkably close
to the 1-loop string unification scale.

Overall we find that the idea of SPU is both motivated and 
potentially testable at the next generation of colliders
through its novel effects on the sparticle spectrum.

\section*{Acknowledgements}
We would like to thank K.~Dienes,
J.~Ellis, L.~Hall, I.~Jack, T.~Jones, H.~Murayama,
G.~Ross and especially K.S.~Babu for useful conversations.
We would both like to acknowledge 
the hospitality of the Aspen Center for Physics,
and one of us (JMR) the hospitality of the Lawrence Berkeley National
Laboratory, where parts of this work were performed.

\section*{Appendix}

The form of the 3-loop $\be$-functions is
\beq
{d\al_i\over dt} = \al_i^2\left\lbrace\frac{b_i}{2\pi} +
\frac{1}{8\pi^2}\left(b_{ij}\al_j+a_{ia}\ytil_a\right) +
\frac{1}{32\pi^3}\left(b_{ijk}\al_j\al_k + c_{ija}\al_j\ytil_a +
a_{iab}\ytil_a\ytil_b\right)\right\rbrace
\label{eq:rgeform}
\eeq
where $\ytil_a=y_a^2/4\pi$.
In the case of an additional $\nf$ $\five$'s and $\nt$
$\ten$'s, the 1- and 2-loop coefficients are well-known:
\beq
b_i = [33/5,1,-3] +\frac{1}{2}(\nf+3\nt)[1,1,1]
\label{eq:bi}
\eeq
and
\beq
b_{ij} = \left[ \begin{array}{ccc}
\fract{199}{25} & \fract{27}{5} & \fract{88}{5} \\
\fract{9}{5}  & 25 & 24\\
\fract{11}{5}  & 9 &  14
\end{array} \right] +
\nf \left[ \begin{array}{ccc}
\fract{7}{30}  & \fract{9}{10} & \fract{16}{15} \\
\fract{3}{10}  & \fract{7}{2} & 0 \\
\fract{2}{15} & 0 &  \fract{17}{3}
\end{array} \right] +
\nt \left[ \begin{array}{ccc}
\fract{23}{10} & \fract{3}{10} & \fract{24}{5} \\
\fract{1}{10} & \fract{21}{2} & 8 \\
\fract{3}{5} & 3 & 17
\end{array} \right]
\eeq
The $a_{ia}$ can be found in the literature and do not change as additional
matter is added.

The 3-loop coefficients in the $\ol{\rm DR}$-scheme
for a simple gauge group have been recently
calculated in
Ref.~\cite{jj1}. This was extended to the MSSM in Ref.~\cite{jj2};
expressions for the $a_{iab}$, $b_{ijk}$ and
$c_{ija}$ in the MSSM (or in the MSSM with extra $\sixteen$'s)
can be easily extracted from the explicit expressions
given there. For the purposes of this study we are setting the
Yukawa contributions
in the 3-loop RGE's to zero, keeping only the pure gauge pieces.

That there is a scheme dependence to the coefficients of the 3-loop
$\beta$-functions is well-known. Our choice of $\ol{\rm DR}$ however
is the natural one, since it is within this scheme that gauge and gaugino
unification are expected to hold.

The 3-loop gauge contributions to the RGE's for a product group
can be written in the $\overline{\rm DR}$ scheme as:
\beq
\left.\frac{d\al_i}{dt}\right|_{\rm 3-loop}&=&
\frac{\al_i^2}{32\pi^3}\left\lbrace\al_i^2
b_iC(G_i)[4C(G_i)-b_i] + 8C(G_i)\sum_{a,j} \al_i\al_jT_i(R_a)C_j(R_a)
\right. \nonumber \\ & & \left.
-{}6\sum_{a,j}\al_j^2b_jT_i(R_a)C_j(R_a)-8\sum_{a,j,k}\al_j\al_kT_i(R_a)
C_j(R_a)C_k(R_a)\right\rbrace
\eeq
where $i,j,k$ label gauge groups and $a$ labels matter representations.
The Casimirs have the usual definitions:
\beq
C_i(R)\delta_m^n\equiv({\bf t}_i^A{\bf t}_i^A)_m^n\quad\quad\quad\quad
T_i(R)\delta^{AB}\equiv {\rm Tr}_R({\bf t}_i^A{\bf t}_i^B)
\eeq
with ${\bf t}_i$ the generators of gauge group $i$. In our normalization,
for $SU(N)$: $C(G)=N$, $T(R)=\frac{1}{2}$ for a fundamental, and $C(R)=
\frac{3}{4}$ or $\frac{4}{3}$ for a fundamental of $SU(2)$ or $SU(3)$.

Plugging into the general form for the MSSM with additional
$\five$'s and $\ten$'s, one finds
the $b_{ijk}$ to be:
\bea
b_{1jk}\al_j\al_k&=& \left(-\fracs{32117}{375}-\fracs{7507}{900}n_5
-\fracs{12859}{300}n_{10}-\fracs{7}{40}n_5^2-\fracs{207}{40}n_{10}^2
-\fracs{9}{4}n_5n_{10}\right)\al_1^2 \nonumber \\ & &
+{}\left(-\fracs{81}{5}-\fracs{27}{4}n_5-\fracs{261}{20}n_{10}
-\fracs{27}{40}n_5^2
-\fracs{27}{40}n_{10}^2-\fracs{9}{4}n_5n_{10}\right)\al_2^2 \nonumber \\ & &
+{}\left(\fracs{484}{15}-\fracs{506}{45}n_5-\fracs{154}{5}n_{10}
-\fracs{4}{5}n_5^2
-\fracs{54}{5}n_{10}^2-6n_5n_{10}\right)\al_3^2 \nonumber \\ & &
+{}\left(-\fracs{69}{25}-\fracs{27}{50}n_5-\fracs{1}{50}n_{10}\right)\al_1\al_2
+\left(-\fracs{1096}{75}-\fracs{64}{225}n_5-\fracs{344}{75}n_{10}\right)
\al_1\al_3 \nonumber \\ & &
+{}\left(-\fracs{24}{5}-\fracs{8}{5}n_{10}\right)\al_2\al_3,
\nonumber \\
b_{2jk}\al_j\al_k&=& \left(-\fracs{457}{25}-\fracs{441}{100}n_5
-\fracs{1513}{300}
n_{10}-\fracs{9}{40}n_5^2-\fracs{9}{40}n_{10}^2
-\fracs{3}{4}n_5n_{10}\right)\al_1^2 \nonumber \\ & &
+{}\left(35-\fracs{33}{4}n_5-\fracs{99}{4}n_{10}-\fracs{13}{8}n_5^2
-\fracs{117}{8}
n_{10}^2-\fracs{39}{4}n_5n_{10}\right)\al_2^2 \nonumber \\ & &
+{}\left(44-18n_5-\fracs{118}{3}n_{10}-18n_{10}^2-6n_5n_{10}\right)\al_3^2
\\ & &
+{}\left(\fracs{9}{5}+\fracs{3}{10}n_5+\fracs{1}{10}n_{10}\right)\al_1\al_2
+\left(-\fracs{8}{5}-\fracs{8}{15}n_{10}\right)\al_1\al_3 \nonumber \\ & &
+{}\left(24+8n_{10}\right)\al_2\al_3,
\nonumber \\
b_{3jk}\al_j\al_k&=& \left(-\fracs{1702}{75}-\fracs{2689}{900}n_5-\fracs{3353}
{300}n_{10}-\fracs{1}{10}n_5^2-\fracs{27}{20}n_{10}^2-\fracs{3}{4}n_5n_{10}
\right)\al_1^2 \nonumber \\ & &
+{}\left(-27-\fracs{27}{4}n_5-\fracs{117}{4}n_{10}-\fracs{27}{4}n_{10}^2
-\fracs{9}{4}n_5n_{10}\right)\al_2^2 \nonumber \\ & &
+{}\left(\fracs{347}{3}+\fracs{215}{9}n_5+\fracs{215}{3}n_{10}
-\fracs{11}{4}n_5^2
-\fracs{99}{4}n_{10}^2-\fracs{33}{2}n_5n_{10}\right)\al_3^2 \nonumber \\ & &
+{}\left(-\fracs{3}{5}-\fracs{1}{5}n_{10}\right)\al_1\al_2
+\left(\fracs{22}{15}+\fracs{4}{45}n_5+\fracs{2}{5}n_{10}\right)\al_1\al_3
\nonumber \\ & &
+{}\left(6+2n_{10}\right)\al_2\al_3. \nonumber
\eea
The MSSM is recovered for $n_5=n_{10}=0$.

\end{document}